\documentclass[prd,aps,lengthcheck,twocolumn,nofootinbib,notitlepage,floatfix,superscriptaddress]{revtex4-2}
\usepackage{graphics,graphicx}
\usepackage{amsmath, amssymb}
\usepackage{multirow}
\usepackage{color}
\usepackage{verbatim}
\usepackage{hyperref}
\usepackage[normalem]{ulem}
\usepackage{color}
\usepackage{cancel}
\usepackage{mathtools}

\def\fm3{\;\text{fm}^{-3}}

\renewcommand\sout{\bgroup\color{blue} \ULdepth=-.5ex \ULset}
\def\slashchar#1{\setbox0=\hbox{$#1$}  
\dimen0=\wd0     
\setbox1=\hbox{/} \dimen1=\wd1  
\ifdim\dimen0>\dimen1   
\rlap{\hbox to \dimen0{\hfil/\hfil}} 
#1     
\else     
\rlap{\hbox to \dimen1{\hfil$#1$\hfil}} 
/      
\fi}

\newcommand{\eps}{\epsilon}
\newcommand{\dd}{\mathrm{d}}

\begin{document}
\title{A string based model with Hagedorn temperature of $T_H\sim 300~$MeV describes the spectrum of mesons and glueballs}

\author{Micha\l{} Marczenko}
\email{michal.marczenko@uwr.edu.pl}
\affiliation{Incubator of Scientific Excellence - Centre for Simulations of Superdense Fluids, University of Wroc\l{}aw, 
plac Maksa Borna 9, 50-204 Wroc\l{}aw, Poland}
\affiliation{Institute of Theoretical Physics, University of Wroc\l{}aw, 
plac Maksa Borna 9, 50-204 Wroc\l{}aw, Poland}
\author{Gy\H{o}z\H{o} Kov\'acs}
\email{gyozo.kovacs@uwr.edu.pl}
\affiliation{Institute of Theoretical Physics, University of Wroc\l{}aw, 
plac Maksa Borna 9, 50-204 Wroc\l{}aw, Poland}
\affiliation{Institute for Particle and Nuclear Physics, HUN-REN Wigner Research Centre for Physics, 1121 Budapest, Konkoly–Thege Miklós út 29-33, Hungary}
\author{Larry McLerran}
\affiliation{Institute for Nuclear Theory, University of Washington, Box 351550, Seattle, WA 98195, USA}
\author{Krzysztof Redlich}
\affiliation{Institute of Theoretical Physics, University of Wroc\l{}aw, 
plac Maksa Borna 9, 50-204 Wroc\l{}aw, Poland}
\affiliation{Polish Academy of Sciences PAN, Podwale 75, 
50-449 Wroc\l{}aw, Poland}

\date{\today}
\begin{abstract}
We consider the thermodynamics of a color-confined phase of quantum chromodynamics (QCD) and pure gauge theory within a string-inspired model, corresponding to a physical spatial dimension, d = 3. We show that the physical mass spectrum of massive mesons--in both the strange and non-strange sectors separately--is reasonably well described and extended by the exponential mass spectrum of open strings, $\rho(m)$, characterized by a unique Hagedorn temperature, $T_H = \sqrt{3\sigma/2\pi}$, expressed by the string tension, $\sigma$. This $T_H$ is the value appropriate for d = 3 spatial dimensions, and is of order $T_H \sim 300~\rm MeV$ for typical values of the string tension.  It is much larger than the values of $T_H$, which have been phenomenologically extracted
so far to describe the meson spectrum. Glueball states in pure gauge theory, modeled by closed strings, exhibit a similarly large Hagedorn temperature, highlighting a universal feature of the exponential spectrum. We further analyze the thermodynamic properties of the equation of state at finite temperature and demonstrate that, in the confined phase, the string models agree with lattice QCD results. This lends further support to the recent interpretation of the QCD phase diagram that incorporates strings as relevant degrees of freedom.
\end{abstract}

\maketitle

\section{Introduction}
\label{sec:introduction}

The determination of the phase diagram of quantum chromodynamics (QCD) is one of the most interesting topics in high-energy physics,~see, e.g.,~\cite{Fukushima:2013rx, Gross:2022hyw, Karsch:2022jwp, Du:2024wjm} for a recent theoretical overview. At finite temperature and low net-baryon number density, studies of QCD on the lattice (LQCD) were crucial in arguing for the existence of the transition from hadronic matter to a quark-gluon plasma (QGP)~\cite{Bazavov:2014pvz, Borsanyi:2018grb, Bazavov:2017dus, Bazavov:2020bjn, Bazavov:2020bjn, Aoki:2006we}. The nature of the transition is a smooth crossover that occurs at the pseudocritical temperature, $T_c \simeq 156~$MeV~\cite{HotQCD:2018pds}. 

Recently, a novel state of matter was proposed~\cite{Cohen:2023hbq, Fujimoto:2025sxx} to exist between the confined hadronic phase and deconfined QGP, i.e., at temperatures between the chiral crossover, $ T_c\simeq 156~$MeV, and the separated deconfinement temperature $T_d$. In Ref.~\cite{Fujimoto:2025sxx}, the so-called {\it Spaghetti of Quarks with Glueballs} (SQGB) phase was introduced with an upper boundary at $T_d \simeq 300~$MeV corresponding to a pure gauge theory~\cite{Meyer:2009tq, Athenodorou:2020ani, Borsanyi:2022xml, Giusti:2016iqr, Giusti:2025fxu}. 

Phenomenologically, the appearance of deconfinement in QCD matter at finite temperature could be potentially linked with the concept of the Hagedorn limiting temperature, $T_H$, which is the upper limit of permissible temperature of hadronic matter. The concept of $T_H$ was first introduced by Hagedorn~\cite{Hagedorn:1965st} within the Statistical Bootstrap Model (SBM)
to explain the rapid growth of the experimentally observed hadronic spectrum, providing a self-consistent statistical description in which hadrons are composed of other hadrons in a self-similar way~\cite{Hagedorn:1971mc, Frautschi:1971ij, Rafelski:2016nxx}. In the SBM
the hadron mass spectrum is exponential, $\rho(m) \simeq m^a e^{m/T_H}$, with $a$ and $T_H$ being model parameters. Formulating the thermodynamics for such $\rho(m)$ implies that the power law coefficient $a$ determines the divergence of different thermal observables at $T_H$  \cite{Hagedorn:1971mc}. This is what made Hagedorn conclude that $T_H$ is the highest possible temperature of hadronic matter.

The value of $T_H$, extracted from different implementations of the bootstrap conditions \cite{Hagedorn:1971mc, Rafelski:2016nxx} or from the recent fit of $\rho(m)$ to the Particle Data Group (PDG) data ~\cite{ParticleDataGroup:2024cfk}, as well from matching $\rho(m)$ thermodynamics to LQCD results  \cite{Majumder:2010ik,Lo:2015cca, ManLo:2016pgd, Broniowski:2000bj, Lo:2015cca, Broniowski:2004yh}, lies in the range, $135< T_H<190$ MeV, depending on the value and the form of the preexponential factor. This range contains the chiral crossover temperature $T_c\simeq 156$~MeV but is essentially lower than deconfinement temperature $T_d\simeq 300$~MeV from pure gauge theory~\cite{Meyer:2009tq, Athenodorou:2020ani, Borsanyi:2022xml, Giusti:2016iqr, Giusti:2025fxu}.

The exponential Hagedorn spectrum also emerges naturally in models where hadrons are described as excitations of a relativistic string~\cite{Green_Schwarz_Witten_2012}. An exponential growth of the density of states characterizes it. In this framework, the exponential proliferation of states is a direct consequence of the large number of accessible oscillator modes at high excitation levels. The resulting Hagedorn temperature $T_H$ defines the limiting scale in thermodynamics. 
Its value is directly linked to the string tension, $\sqrt{\sigma} \simeq 440~$MeV~\cite{Meyer:2004gx}, which in three-dimensional theory yields a significantly higher Hagedorn temperature, $T_H =\sqrt{3\sigma /2\pi} \simeq 300~$MeV~\cite{Fujimoto:2025sxx}, reflecting the melting of long flux tubes in pure gauge theory. We note that the underlying string theory provides also a precise determination of the preexponential factor.

One of the arguments behind the new phase proposed in~\cite{Fujimoto:2025sxx} is based on the assumption that the confined phase of QCD and pure gauge theory have string representations. The mass spectrum of mesons and glueballs is exponential and described respectively by the spectrum of open and closed strings with a common and large Hagedorn temperature linked to the string tension. Consequently, a new concept of the QCD phase diagram introduced in Ref.~\cite{ Fujimoto:2025sxx} hinges on the fact that the Hagedorn limiting temperature can be related to the deconfinement temperature in pure gauge theory, $T_H \simeq T_d\sim 300$ MeV. This highlights the important distinction between the interpretation of $T_H$ as a breakdown of the hadronic description and the deconfinement of fundamental color degrees of freedom.

The main objective of this paper is to demonstrate that a three-dimensional string theory formulated for open and closed strings provides a complete determination of the meson and glueball spectra with the same Hagedorn temperature of the order of $T_H \simeq300~$MeV, albeit different but uniquely described preexponential factors. We also show that the thermodynamics formulated with the asymptotic string-type Hagedorn mass spectrum  is consistent with LQCD results on the equation of state (EoS) in QCD and pure gauge theory.

The paper is organized as follows. In Sec.~\ref{sec:had_eos_discrete}, we introduce the HRG model with a discrete mass spectrum. In Sec.~\ref{sec:hag_mass}, we introduce the Hagedorn mass spectrum for glueballs and mesons. In Sec.~\ref{sec:thermo}, we discuss the equation of state in conjunction with LQCD findings. Finally, Sec.~\ref{sec:sum} is devoted to summary and conclusions.

\section{Discrete Mass spectrum}
\label{sec:had_eos_discrete}

A phenomenological description of hadronic matter requires identifying the relevant degrees of freedom and their interactions. In the confined phase of QCD, the medium is composed of hadrons and their resonances. The commonly used model to account for the hadronic interactions at low energy is the hadron resonance gas (HRG) model
~\cite{Braun-Munzinger:2003pwq,Andronic:2017pug}. In its simplest form, the decay properties of the resonances are neglected; thus, they are treated as stable point-like particles. The medium composition in the HRG model is given via the density of states as follows
\begin{equation}\label{eq:rho_pdg}
    \rho(m) = \sum_i d_i \delta \left(m - m_i \right) \textrm,
\end{equation}
where the sum goes through all stable hadrons and resonances with mass $m_i$ and spin degeneracy $d_i$. 

It is customary to consider the cumulative mass spectrum~\cite{Broniowski:2000bj}
\begin{equation}\label{eq:cum_def}
    N(m) = \int\limits_0^m \dd m'\; \rho(m') \textrm,
\end{equation}
such that
\begin{equation}
    \rho(m) = \frac{\dd N(m')}{\dd m'}\Bigg|_{m'=m}
\end{equation}
Thus, $N(m)$ counts the number of degrees of freedom with mass below $m$. For the density of states in Eq.~\eqref{eq:rho_pdg} one gets
\begin{equation}
    N(m) = \sum_i d_i\theta\left(m-m_i\right) \textrm,
\end{equation}
where $\theta$ is the Heaviside step function.

Given the additive structure of the density of states in Eq.~\eqref{eq:rho_pdg}, it can be decomposed into contributions from mesons and baryons, as well as definite quantum numbers~\cite{Lo:2015cca}.

In this work, we include established mesons and baryons with three- and four-star ratings from the PDG. We note that $f(500)$ and $\kappa^\star_0(700)$ mesons are not included in the spectra due to their ambiguous nature~\cite{Broniowski:2015oha, Friman:2015zua}. The spectrum of experimentally established hadrons accounts for all identified states with masses up to $m\simeq 2.4~$GeV for mesons and $m\simeq 2.6~$GeV for baryons. The complicated decay properties and large width of the resonances make the investigation of higher excited states challenging. 

\begin{figure}[t!]
    \centering
    \includegraphics[width=1\linewidth]{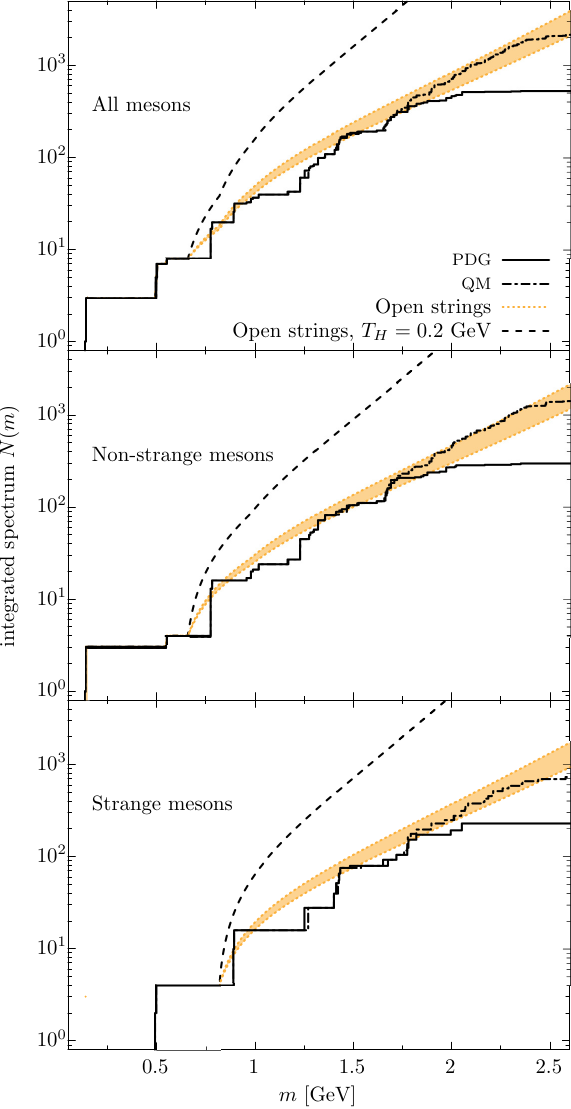}
    \caption{Cumulative mass spectra of all (top), non-strange (middle), and strange (bottom) mesons in the PDG (black, solid lines). Also shown are spectra that include the prediction from the quark model~\cite{Loring:2001ky, Ebert:2009ub} (QM) (black, dash-dotted lines). The yellow bands represent the uncertainty in the Hagedorn limiting temperature in the exponential spectrum (see text for details). We note that $f(500)$ and $\kappa^\star_0(700)$ mesons are not included in the discrete spectra due to their ambiguous nature~\cite{Broniowski:2015oha, Friman:2015zua}. The black, dashed lines show spectra obtained for $T_H = 0.2~$GeV.}
    \label{fig:spectra_meson}
\end{figure}

In Fig.~\ref{fig:spectra_meson}, we show the cumulative mass spectrum of mesons in different sectors of the strange quantum number with inputs from PDG. It is shown for all mesons in the top panel of the figure. For masses up to $m\lesssim 2~$GeV, the increase of the density of states is roughly linear on the logarithmic scale. A similar trend is observed for non-strange and strange mesons, shown in the middle and bottom panels, respectively. We also verified it for the list of resonances that includes the prediction from the quark model (QM)~\cite{Loring:2001ky, Ebert:2009ub}, where the linear trend continues up to $m\simeq 2.5~$GeV. This indicates that the meson mass spectrum is exponential and can be described by the Hagedorn mass spectrum~\cite{Hagedorn:1965st, Hagedorn:1971mc}.

Analogous to the HRG model, the glueball resonance gas (GRG) model describes the confined phase of pure Yang-Mills theory as a non-interacting gas of glueballs and their excitations. The glueball spectrum has been evaluated in several studies using LQCD methods (see, e.g.,~\cite{Meyer:2004gx, Chen:2005mg, Athenodorou:2020ani, Chen:2004bw}). We note that the differences in the values of the string tension obtained in LQCD simulations result in a non-negligible shift in the physical scale of all dimensionful observables, such as glueball masses and critical temperature~\cite{Trotti:2022knd}. In this work, we use the spectra from Refs.~\cite{Meyer:2004gx} and~\cite{Athenodorou:2020ani} which obtain the string tension $\sqrt\sigma = 440(20)~$MeV and $\sqrt\sigma = 485(6)~$MeV, respectively. This discrepancy reflects the inherent ambiguity in setting an absolute scale in pure gauge theory, where no physical input exists to fix the units uniquely. For this reason,  expressing results in units of the string tension $\sqrt{\sigma}$ is a robust and universal approach that eliminates uncertainties related to the physical scale and facilitates comparison between different LQCD simulations and phenomenological models.

\begin{figure}[t!]
    \centering
    \includegraphics[width=1\linewidth]{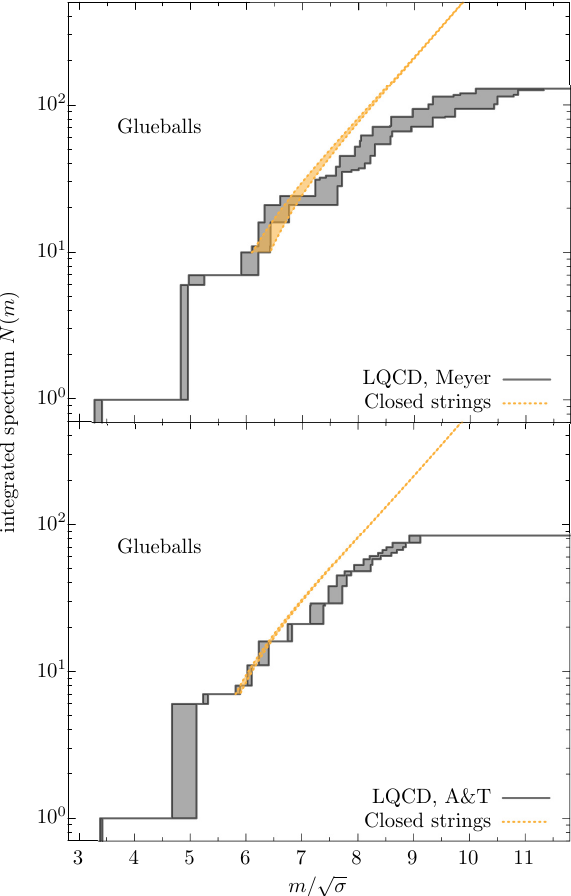}
    \caption{Continuum-extrapolated cumulative mass spectra of glueballs from LQCD simulations (black, solid bands). The spectra are taken from Ref.~\cite{Meyer:2004gx} (top panel) and~\cite{Athenodorou:2020ani} (bottom panel) and are depicted in the units of the string tension $\sqrt\sigma$. The bands represent the uncertainties of the continuum-limit extrapolation. The spectra for closed strings are shown as orange, dashed bands. Their uncertainties come from the uncertainties of the continuum-limit extrapolation of the mass of the lightest resonance in the LQCD spectra (see text).}
    \label{fig:spectrum_gb}
\end{figure}

In Fig.~\ref{fig:spectrum_gb}, we present the continuum-extrapolated cumulative mass spectrum for glueballs obtained in the $SU(3)$ pure-gauge LQCD simulations~\cite{Meyer:2004gx, Athenodorou:2020ani}. The number of states grows approximately exponentially with mass, consistent with a Hagedorn-like spectrum. However, a precise extraction of the exponential slope is difficult due to limited statistics and uncertainties in spin identification at higher masses. The spectra from Refs.~\cite{Meyer:2004gx, Athenodorou:2020ani} are broadly consistent in state count for $SU(3)$, though they differ in methodology and the treatment of higher excitations.

The observed exponential trend of the experimental and theoretical spectra for mesons and glueballs indicates the exponential mass spectrum predicted by Hagedorn in the context of the Statistical Bootstrap Model~\cite{Hagedorn:1965st, Hagedorn:1971mc} and then found in a dual string and bag models~\cite{Huang:1970iq, Cudell:1992bi, Johnson:1975sg}.

\section{Hagedorn mass spectrum}
\label{sec:hag_mass}

From Figs.~\ref{fig:spectra_meson} and~\ref{fig:spectrum_gb} it is clear that for larger masses the cumulative mass spectrum becomes better and better approximated by a smooth function of $m$ since new states appear more frequently and the number of these states are relatively smaller compared to the cumulated number. In this high mass region, the spectrum can be parameterized by the Hagedorn exponential form, $\rho^H(m)\simeq m^ae^{m/T_H}$.

To  describe  the spectrum, we adopt the idea of treating the ground-state particles separately from the continuous spectrum. In practice, one uses the cutoff scale $m_x$, such that the continuous spectrum starts at the mass of the lightest resonance in the spectrum. Therefore, we apply the following mass spectrum
\begin{equation}\label{eq:disc_had}
    \rho(m) = \sum_{i}\delta(m-m_i) + \theta(m-m_x) \rho^H(m) \textrm,
\end{equation}
where the index $i$ counts states with mass below $m_x$. The cumulative mass spectrum can be computed straightforwardly from Eq.~\eqref{eq:cum_def}. 

In the string model, a Hagedorn spectrum is generated.  In the discussion below, we will consider such a string theory description of $\rho^H(m)$ for physical spatial dimension $d = 3$. This string-based description predicts a universal Hagedorn temperature for mesons and glueballs in terms of the string tension.  The pre-factor of the exponential spectrum is also uniquely determined and is dependent upon whether one considers mesons (open strings) or glueballs (closed strings).  The meson spectrum may be decomposed in terms of the quantum numbers of mesons, with appropriate multiplicative degeneracy factors for spin and flavor.

For the string model in physical spatial dimensions $d=3$, hence $d_\perp=2$ in the exponential formula, there are profound difficulties with computing the low-lying mass spectra--associated with ghosts and tachyons that are absent only for $d=26$--and properly describing pseudo-Goldstone bosons. Here, we will assume that the string theory describes only the higher mass excitations, just as in the case of the usual Hagedorn spectrum, hence we do not need to consider these issues.

\subsection{Glueballs}

Following Ref.~\cite{Fujimoto:2025sxx}, we model the exponential glueball spectrum by the spectrum of closed strings
\begin{equation}\label{eq:closed_spec}
    \rho^H_{\rm gb}(m) = \frac{1}{T_H}\left(\frac{2\pi}{3}\right)^3\left(\frac{T_H}{m}\right)^4 e^{m/T_H} \textrm,
\end{equation}
with $T_H$ being the Hagedorn limiting temperature, which is related to the string tension $\sqrt{\sigma}$ as
\begin{equation}\label{eq:th_sigma}
    T_H = \sqrt{\frac{3}{2\pi}}\sqrt{\sigma}  \textrm.
\end{equation}
In the current study, we take the threshold mass $m_x$ to be the mass of the lightest glueball resonance in the spectrum. 

In Fig.~\ref{fig:spectrum_gb}, we show the cumulative glueball spectra from $SU(3)$ pure-gauge LQCD simulations. We note that in the units of $\sqrt\sigma$, the Hagedorn temperature $T_H =  \sqrt{{3}/{2\pi}} \approx 0.691$ and the only source of uncertainties comes solely from the uncertainty in the threshold mass $m_x$. We find a surprising agreement of the continuous Hagedorn spectrum up to $m/\sqrt\sigma\simeq 7$ for the spectra from~\cite{Meyer:2004gx} and~\cite{Athenodorou:2020ani}. This may point to hitherto unknown glueball resonances in the spectrum from LQCD simulations. Interestingly, the deviation starts roughly at the two glueball mass thresholds. We note that determining the glueball spectrum from LQCD is a technically challenging task, and the complete and unambiguous spectrum in full QCD remains difficult to resolve due to operator mixing, statistical noise, and lattice artifacts. Because of these difficulties, the calculable lattice spectrum is limited and begins to saturate with missing high-mass states before it can reach a sufficiently large $m/\sqrt{\sigma}$ to exhibit a smooth exponential behavior. However, as it seems from the thermodynamics, presented in a subsequent section,  the states, which are probably missing from the LQCD spectrum above $m/\sqrt{\sigma}\approx 7$, but present in the Hagedorn spectrum of closed strings, are important for reproducing the lattice QCD results on the thermodynamic quantities.

\begin{figure}[t!]
    \centering
    \includegraphics[width=1\linewidth]{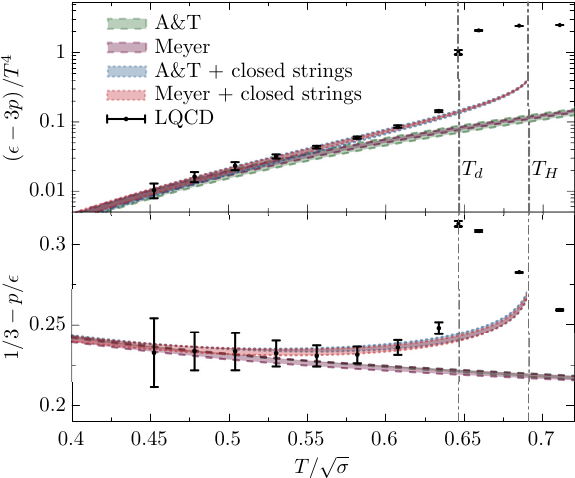}
    \caption{Trace anomaly $(\eps - 3p)/T^4$ (top panel) and energy-density-normalized trace anomaly $1/3-p/\eps$ (bottom panel). The SU(3) pure gauge LQCD data on pressure and energy density are taken from Ref.~\cite{Borsanyi:2012ve}. The uncertainty bands in both panels are obtained by propagating the reported error on pressure and energy density. The vertical lines $T_d/\sqrt{\sigma}=0.646$ and $T_H/\sqrt{\sigma}=0.691$ represent the critical deconfinement and Hagedorn temperatures, respectively. Note that the closed strings are shown only up to $T_H$ (see text for details).}
    \label{fig:thermo_glue}
\end{figure}

\subsection{Mesons}

Following Ref.~\cite{Fujimoto:2025sxx}, we model the Hagedorn spectrum of mesons by open strings:
\begin{equation}\label{eq:open_spec}
    \rho^H_{\rm mes}(m) = d_\sigma(m)\frac{\sqrt{2\pi}}{6 T_H} \left( \frac{T_H}{m} \right)^{3/2} e^{m/T_H} \textrm,
\end{equation}
with the same Hagedorn temperature $T_H = \sqrt{3\sigma/2\pi}$ as for glueballs.

In pure Yang-Mills theory, the string tension $\sqrt\sigma$ defines the intrinsic physical scale and is extracted from the long-distance behavior of the static quark-antiquark potential. When applying a Hagedorn spectrum to mesons, whose masses are expressed in physical units, it becomes necessary to assign a value to $\sqrt\sigma$ in order to convert theoretical predictions into physical temperatures. Although $\sqrt\sigma$ is not a directly measurable quantity in full QCD due to string breaking, it remains a meaningful and widely used scale, with typical values ranging from $\sqrt\sigma = 0.440(20)~$GeV~\cite{Meyer:2004gx} to $\sqrt\sigma = 0.485(9)~$GeV~\cite{Athenodorou:2020ani}. Here, we use both central values to translate the Hagedorn temperature into physical units and interpret the resulting variation as a systematic uncertainty in the scale setting. Accordingly, we consider the range $T_H = 0.304-0.335~$GeV, where the range reflects the uncertainty associated with $\sqrt\sigma=0.440 - 0.485~$GeV.

The degeneracy factor $d_\sigma(m)$ in Eq.~\eqref{eq:open_spec} varies depending on the mass thresholds
\begin{align}
\label{eq:dsigma}
    d_\sigma(m) =
    \begin{cases}
        16 & {\rm {for }}~~2m_l \le m < m_l + m_s\textrm,\\
        32 & {\rm {for }}~~m_l + m_s \le m < 2m_s\textrm,\\
        36 & {\rm {for }}~~2 m_s \leq m \textrm,
    \end{cases}
\end{align}
where $m_l \simeq 0.33~$GeV and $m_s \simeq 0.49~$GeV are the constituent masses for light and strange quarks, respectively.

The spectrum $\rho(m)$ is additive. Therefore, one may consider the spectra of non-strange ($ns$) and strange ($s$) mesons separately,
\begin{equation}
    \rho_{\rm mes} = \rho^{ns}_{\rm mes} + \rho^{s}_{\rm mes} \textrm.
\end{equation}
The degeneracy factors in $\rho^{ns}_{\rm mes}$ and $\rho^{s}_{\rm mes}$ have to be adjusted to count the non-strange and strange degrees of freedom separately, namely
\begin{align}
\label{eq:dsigma_s0}
    d^{ns}_\sigma(m) =
    \begin{cases}
        16 & {\rm {for }}~~2m_l \le m < 2m_s\textrm,\\
        20 & {\rm {for }}~~2 m_s \leq m  \textrm,
    \end{cases}
\end{align}
for the non-strange mesons, and
\begin{align}
\label{eq:dsigma_s1}
    d^{s}_\sigma(m) = 16~{\rm {for }}~m \geq m_l+ m_s  \textrm,
\end{align}
for the strange mesons. The mass thresholds are given as $m_x^{ns} = 2m_l = 0.67~$GeV and $m_x^{s} = m_l+m_s = 0.82~$GeV. The discrete part of the spectrum includes the Nambu-Goldstone mesons, i.e., $\lbrace\pi,~K,~\eta\rbrace$ below the threshold masses.

The continuous exponential Hagedorn spectra for mesons are shown in Fig.~\ref{fig:spectra_meson}. The cumulative spectra of all mesons, as well as the separate subsets of strange and non-strange mesons, are well described by a common exponential Hagedorn form. While the strange meson spectrum is somewhat less dense, the overall agreement remains compelling and consistent with a high Hagedorn limiting temperature $T_H$. These results support the idea that the exponential growth in the hadronic spectrum with $T_H\simeq 300~$MeV persists across different quantum number sectors if the prefactors $d_\sigma(M)$ are chosen correctly. There remains, however, some ambiguity in choosing the precise mass range where the exponential behavior becomes dominant, since the low-mass part of the spectrum is dominated by well-known resonances that are sparse and do not yet follow the exponential trend. Nevertheless, the high-mass part of the spectrum is rather insensitive to modest shifts of the starting point, and the $T_H\simeq 300~$MeV is robust.

\begin{figure}[t!]
    \centering
    \includegraphics[width=1\linewidth]{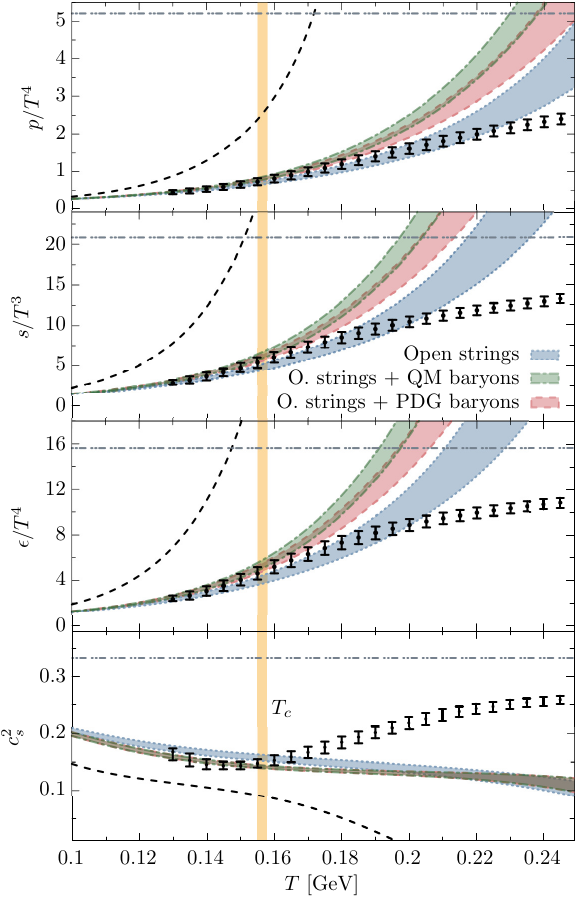}
    \caption{The equation of state at finite temperature and vanishing chemical potential. The LQCD results are taken from Ref.~\cite{HotQCD:2014kol}. Mesons are modeled via the continuous spectrum of open strings (cf.~Eq.~\eqref{eq:open_spec}), and baryons are taken as discrete states from the Particle Data Group (PDG) or quark model (QM). The blue, red, and green bands indicate the uncertainty in the string tension (see text). The yellow vertical band marks the estimation of the pseudocritical temperature for the chiral crossover transition $T_c=156.5\pm1.5~$MeV~\cite{HotQCD:2018pds}. The gray, horizontal, doubly-dotted--dashed lines mark the Stefan-Boltzmann limit considering quarks of 3 flavors and gluons. The black, dashed lines show results for open strings with $T_H = 0.2~$GeV.}
    \label{fig:thermo}
\end{figure}

We find a remarkably good agreement between the cumulative meson spectrum and the exponential form with a high Hagedorn temperature, expected from string-theoretic considerations. This temperature is significantly higher than the one typically associated with the breakdown of the hadron resonance gas model ($T_H \lesssim 0.2$~GeV). The large common $T_H$ for different types of hadrons implies that a large number of hadronic states can remain intact and thermodynamically relevant well beyond the conventional crossover temperature. This can apply particularly for the glueballs, since the mesons become unfavored for the thermodynamic description, as shown in the subsequent section.

\section{Thermodynamics}
\label{sec:thermo}

To investigate the thermodynamic behavior of the hadronic phase, we model the system as a gas of non-interacting hadronic states governed by an exponential Hagedorn spectrum. At finite temperature $T$ and vanishing chemical potential, the pressure of a noninteracting relativistic particle species with mass $m$ in the Boltzmann approximation reads
\begin{equation}
    p(m) = \frac{m^2T^2}{2\pi^2} K_2\left(\frac{m}{T}\right) \textrm,
\end{equation}
where $K_2$ is the modified Bessel function of the second kind. Given the density of states $\rho(m)$, the total pressure can be written as
\begin{equation}
    p = \int\dd m\; \rho(m)p(m) \textrm,
\end{equation}
with $\rho(m)$ given by Eq.~\eqref{eq:disc_had}. The entropy density $s$, energy density $\epsilon$, and the sound velocity $c_s^2$ can be derived from the pressure through the thermodynamic identities, $s = \dd p / \dd T$, $\eps = Ts - p$, and $c_s^2 = \dd p / \dd \eps$, respectively.

\subsection{Yang-Mills equation of state}
\label{sec:eos_gb}

In Fig.~\ref{fig:thermo_glue}, we show the pure-gauge equation of state at finite temperature in the units of string tension, $T/\sqrt\sigma$. In the top panel of the figure, we consider the trace anomaly $\Delta = (\eps - 3p)/T^4$ and energy-density-normalized trace anomaly $\Delta_\eps = \Delta / 3\eps$. The value of the critical temperature of the first-order deconfinement phase transition in pure gauge theory is $T_d/\sqrt\sigma = 0.596(4) + 0.453(30)/N_c^2$~\cite{Lucini:2003zr}. For $N_c = 3$, one gets $T_d/\sqrt\sigma = 0.646(37)$. We note that the Hagedorn temperature equals $T/\sqrt\sigma = \sqrt{3/2\pi} = 0.691$. 
The GRG results deviate from the LQCD data around $T/\sqrt\sigma \simeq 0.57-0.60$. This difference may originate from the growing relevance of heavy glueball resonances, whose contribution is not fully captured in the discrete spectrum. Incorporating an exponential Hagedorn spectrum from closed string model remarkably improves the agreement with the LQCD results. Finally, we observe a clear coincidence between results obtained from different glueball spectra~\cite{Meyer:2004gx, Athenodorou:2020ani}, which highlights the importance of a scale-free comparison.

\subsection{$(2+1)$-flavor QCD equation of state}
\label{sec:eos_had}

Due to the additive nature of the density of states, the hadronic pressure can be decomposed into partial pressures of mesons and baryons
\begin{equation}
    p_{\rm had} = p_{\rm mes} + p_{\rm bar} \textrm,
\end{equation}
respectively.

In Fig.~\ref{fig:thermo}, we show the equation of state at finite temperature and vanishing chemical potential. First, we compare the LQCD EoS to the Hagedorn model with only mesons. The obtained thermodynamic quantities describe the LQCD data up to the pseudocritical temperature. The uncertainty in the Hagedorn temperature, $T_H = 0.304-0.385~$GeV, becomes vivid only at higher temperatures.  

In general, the contribution from baryons should also be taken into account, although their contribution is thermally suppressed. For a fully consistent setup, the baryons could also be built in the string picture. However, even in the simplest scenario, this requires $N_c$ strings joined together at an $N_c$-junction.
The treatment of this case and the derivation of the exponential spectrum are nontrivial and worth a longer discussion. Therefore, we postpone this problem to future work. For simplicity, here we include the contribution of baryons through a discrete density of states as in the ideal HRG model. Including baryons leads to slightly increased thermodynamic quantities ($p$, $s$, and $\epsilon$) that are still consistent with the LQCD EoS up to $T_{c}$. We have checked that the inclusion of the additional baryon resonances predicted by the quark model satisfies the LQCD EoS as well. This is depicted in Fig.~\ref{fig:thermo}. We also note that the contribution of the continuous glueball spectrum to the thermodynamics of the confined phase of QCD with $N_f=2+1$ flavor is negligible~\cite{Fujimoto:2025sxx}. We emphasize, that the open string description--and generally, the mesonic description--is valid only until the $T_{c}$ even if the Hagedorn temperature is much larger, like in our case.

In $(2+1)$-flavor QCD, the speed of sound, $c_s^2$, exhibits a clear minimum around the crossover temperature $T_c$~\cite{HotQCD:2014kol}, emphasizing the possible change of degrees of freedom. The clear deviation of the sound velocity from the LQCD data signals the changeover of the active degrees of freedom in the thermal medium. This also suggests that the agreement between the open string and the LQCD results for $p$, $s$, and $\epsilon$ above $T_{c}$ is likely a coincidence, given their monotonic behaviour.

We note that the normalized trace anomaly $1/3 - p/\eps$ is closely related to the average speed of sound~\cite{Marczenko:2024uit}, thus similarly encodes this behavior and peaks near the critical temperature. This is shown for the pure gauge theory in the bottom panel of Fig.~\ref{fig:thermo_glue}. 

The GRG model exhibits a gradual decrease with temperature, while the inclusion of the exponential Hagedorn spectrum shows an increasing trend, probably due to the relevance of the high mass states for higher $T$. Therefore, it can capture the LQCD results -- where the increase around $T_d$ indicates a rapid change in degrees of freedom and signals the onset of deconfinement.

Thermodynamic quantities involve integrals over the mass spectrum.  For the meson spectrum described by open strings, the entropy and energy density are divergent as $T \rightarrow T_H$, and the pressure is finite.  For the glueball mass spectrum, modeled by closed strings, the entropy, energy density, and pressure are all finite at $T_H$. However, in both cases, the Hagedorn description must break down at temperatures beyond the Hagedorn temperature, since there is a singularity at $T\rightarrow T_H$.

For completeness, we have also examined the model predictions for a smaller Hagedorn temperature, $T_H = 0.2~$GeV. As shown by the dashed curves in Figs.~\ref{fig:spectra_meson} and~\ref{fig:thermo}, such a value overestimates both the cumulative meson spectrum and the lattice thermodynamic observables. The exponential growth of states becomes too steep, leading to a cumulative number of hadrons that exceeds the empirical and quark-model spectra already below $m\simeq1~$GeV. In the thermodynamic sector, the corresponding pressure, entropy density, and energy density rise too rapidly and overshoot the lattice QCD results already at $T\simeq 0.1~$GeV. Furthermore, we note that for glueballs, the Hagedorn temperature cannot be smaller than the deconfinement temperature of pure gauge theory, since the exponential growth of states must terminate at or above the phase transition. Therefore, smaller Hagedorn temperatures typically used in the literature, $T_H \simeq 0.18 - 0.2~$GeV, are inconsistent with both experimental spectra and thermodynamic observables, calculated within the string models considered in this work.

\section{Summary}
\label{sec:sum}
In this work, we have revisited the interpretation of the confined phase of QCD and the SU(3) pure gauge theory using a continuous Hagedorn mass spectrum motivated by the theory of open and closed strings, respectively. In particular, we have shown that a high limiting Hagedorn temperature, $T_H \simeq 300~$MeV--which naturally follows from the string tension within a string model--is consistent with both lattice QCD (LQCD) thermodynamics and experimentally observed mass spectrum.

We demonstrated that the mesonic Hagedorn spectrum derived from the open string model accurately reproduces the experimentally known cumulative meson spectrum, including non-strange and strange mesons separately. This agreement supports the notion that the exponential growth of the hadronic density of states persists across different flavor sectors and that a high Hagedorn temperature is consistent with experimental data.

Furthermore, we showed that the same value of $T_H$ is compatible with the LQCD glueball spectra, modeled as closed-string excitations. While the extraction of higher glueball states using LQCD methods remains challenging, the comparison suggests that the exponential structure of the glueball spectrum is in line with expectations from string theory and does not contradict current LQCD results.

We have also shown that the resulting equation of state, obtained by integrating the Hagedorn spectrum into the hadron resonance gas framework, remains consistent with LQCD data for $N_f = 2 + 1$ QCD up to the chiral crossover temperature.

These findings may lend further support to the proposed existence of an intermediate phase of QCD, the so-called Spaghetti of Quarks with Glueballs (SQGB), where the number of degrees of freedom appears to be that of quarks and glueballs as bound states of gluons~\cite{Fujimoto:2025sxx}. Our results provide phenomenological evidence that a high Hagedorn temperature is not only theoretically motivated but also compatible with the thermodynamic properties of QCD at low intermediate temperatures.

It is vital to assess the effects of the exponential Hagedorn spectrum for baryons. This would allow for further verification of the string picture of hadrons by analyzing, e.g., the fluctuations of conserved charges~\cite{Lo:2015cca}. However, modeling baryons within the string framework is considerably less transparent, as they correspond to more complex string topologies. 

We note that it is remarkable that using a first-principle physics theory, the string, it is possible to successfully describe spectra and thermal properties of the confined phase of (2+1)-flavor QCD and SU(3) pure gauge theory. This is done with no free parameters. This points towards a fundamental feature of the QCD confined phase.

\medskip
\section*{Acknowledgments}

The work of M.~M. was supported through the program Excellence Initiative–Research University of the University of Wroc\l{}aw of the Ministry of Education and Science. 
The work of G.~K. is partially supported by the Polish National Science Centre (NCN) under OPUS Grant No. 2022/45/B/ST2/01527. L.~M.\ thanks partial support from the Institute for Nuclear Theory, which is funded in part by the INT's U.S.\ Department of Energy grant No.\ DE-FG02-00ER41132, and thanks for the support provided by the Physics Department at the University of Wroc\l{}aw where this work was initiated. We also acknowledge fruitful comments from Yuki Fujimoto, Kenji Fukushima, Yoshimasa Hidaka, Bill Zajc and Leonid Glozman. K.~R. thanks Peter Braun-Munzinger, Pok Man Lo, Chihiro Sasaki, and Johanna Stachel, for discussions. We thank Francesco Giacosa for discussions about glueballs.

\bibliography{biblio}
\end{document}